\pdfoutput=1
\documentclass[fleqn,usenatbib]{mnras}

\usepackage{newtxtext,newtxmath}

\usepackage[T1]{fontenc}

\DeclareRobustCommand{\VAN}[3]{#2}
\let\VANthebibliography\thebibliography
\def\thebibliography{\DeclareRobustCommand{\VAN}[3]{##3}\VANthebibliography}

\usepackage{graphicx}	
\usepackage{amsmath}	
\usepackage{amssymb}	

\title[Scattering and exp. discs in \textit{N}-body systems]{Stellar scattering and the formation of exponential discs in self-gravitating systems
 }

\author[J. Wu et al.]{
Jian Wu,$^{1}$\thanks{E-mail: jianwu@iastate.edu (JW)}
Curtis Struck,$^{1}$
Elena D'Onghia$^{2}$
and Bruce G. Elmegreen$^{3}$
\\
$^{1}$Department of Physics and Astronomy, Iowa State University, 2323 Osborn Dr., Ames, IA 50011, USA\\
$^{2}$Department of Astronomy, University of Wisconsin-Madison, 475 N Charter St, Madison, WI 53706, USA\\
$^{3}$IBM Research Division, T.J. Watson Research Centre, 1101 Kitchawan Road, Yorktown Heights, NY 10598, USA
}

\date{Accepted 2020 September 1. Received 2020 August 26; in original form 2020 June 24}

\pubyear{2020}

\hypersetup{draft}

\begin{document}
\label{firstpage}
\pagerange{\pageref{firstpage}--\pageref{lastpage}}
\maketitle

\begin{abstract}
We show, using the \textit{N}-body code 
{\small GADGET}-2, that stellar scattering by massive clumps can produce exponential discs, and the effectiveness of the 
process depends on the mass of scattering centres, as well as the stability of the galactic disc.
Heavy, dense scattering centres in a less stable disc generate an exponential profile quickly, with a
timescale shorter than 1 Gyr. The profile evolution due to scattering can make a near-exponential disc under various initial stellar distributions. This result supports analytic theories that predict the scattering processes always favour the zero entropy gradient solution to the Jeans/Poisson equations, whose profile is a near-exponential. Profile changes are accompanied by disc
thickening, and a power-law increase in stellar velocity dispersion in both vertical and radial directions is also observed through the evolution.
 Close encounters between stars and clumps can produce abrupt changes in stellar orbits and shift stars radially. These events can make trajectories more eccentric, but many leave eccentricities little changed. On average, orbital eccentricities of stars increase moderately with time.

\end{abstract}




\section{Introduction}

Discs with an exponential surface brightness profile are ubiquitous in spiral galaxies, dwarf elliptical galaxies and some irregular galaxies
\citep{boroson1981, ichikawa1986, Patterson1996}.
Near-exponential discs observed in very high redshift galaxies \citep{Elmegreen2005ApJ...634..101E} indicate that an
exponential profile can form fast. Especially, \citet{Bowler2017} showed near-exponential surface brightness profiles for bright Lyman-break galaxies at $z \approx 7$ observed by the Hubble Space Telescope.
Detailed studies on galactic discs show that exponential profiles can be classified into three types \citep{Freeman1970ApJ...160..811F,Pohlen2006,Herrmann2013AJ....146..104H}. A Type I radial profile is a single exponential across the entire galactic disc.
Type II profiles and Type III profiles are composed of an inner exponential and an outer exponential with different slopes. For Type II, the exponential in the outer disc is steeper.
For Type III, the exponential in the inner disc is  steeper.
Observations show that Type II are more common in late-type galaxies \citep{Pohlen2006,Herrmann2013AJ....146..104H}.

Although the observation of exponential profiles is clear, understanding its origin and robustness is difficult.
Popular theories on the formation of an exponential disc include collapse of uniformly rotating spherical
halo without redistribution of angular momentum \citep{Mestel1963}, viscous accretion of gas
\citep{Lin1987}, gravitational influence due to a central bar \citep{hohl1971}, and galaxy interactions \citep{Pe2006ApJ...650L..33P}.
However, as \citet{Elmegreen2013ApJ...775L..35E} pointed out, these theories all have limitations, especially when trying to explain the ubiquity of an exponential disc and its time-efficient formation.
For example, the theory of viscous gas accretion requires shear and differential rotation in 
a disc, but many dwarf galaxies with nearly solid-body rotation establish an exponential.

 \citet{Sellwood2002} showed that spiral arms can scatter stars around the corotation and Lindblad resonance radii, and lead to permanent stellar migration. We refer to it as resonant radial migration. A characteristic of this radial migration is that the change in orbital radius does not significantly increase orbital random energy. Hence a star in a circular orbit remains in a low eccentricity after radial migration. When a bar is present in
 addition to spiral arms, \citet{Minchew2010} showed that the resonance overlap between a bar and spirals can trigger nonlinear responses from stellar orbits. In this scenario, radial mixing of stars takes place in a much faster way.

 Although the theories related to radial migration have been known for  
 many years, the efficiency of this type of mechanism in producing an exponential profile is 
 unclear. On one hand, radial migration can scatter stars from the inner disc to the outer disc, contributing to the creation of an outer exponential \citep{Roskar2008ApJ...675L..65R,sanchez2009MNRAS.398..591S}.
 On the other hand,
 radial migration is more effective when the spiral pattern is strong \citep{Daniel2018MNRAS.476.1561D, Daniel2019ApJ...882..111D} and when 
the stellar population in the disc has low velocity dispersion \citep{Vera-Ciro2014ApJ, Vera-Ciro2016,Daniel2018MNRAS.476.1561D}. 
Observations on high redshift clump-cluster galaxies reveal little evidence on the appearance of spiral patterns \citep{EE2005}, and velocity dispersions in these galaxies are observed to be relatively high \citep{Forster2006ApJ...645.1062F}.

\citet{Elmegreen2013ApJ...775L..35E}  found, with a simplified two-dimensional non-self gravitating model, that an exponential can rise from various initial profiles
due to stellar scattering off from clumps, which represent stellar clusters, giant gas clouds, or other massive objects in a galaxy. 
To allow more complexity, they updated their simulations to three dimensions \citep{Struck2017} and showed that an exponential stellar disc can appear within a few Gyrs and that interstellar holes can scatter stars too. Since massive objects that operate as scattering centres are easy to find in many types of galaxies, the process of this scattering is thought to occur commonly and can be responsible for 
the prevalence of an exponential. In isolated dwarf irregular galaxies, where bars and spiral arms are
weak, this theory may be the only effective channel to achieve an exponential. Among the successes of the updated model, there was still a drawback of non-self-consistency. In their code, stars move as tracers, in the sense that the effect of their motions on gravitational potential is ignored, and clumps are set in fixed circular orbits. One goal of this paper is to break that limitation 
and discuss stellar scattering empowered by self-gravitating simulations. The scattering by holes was softer than that by the clumps \citep{Struck2017}, and produced fewer highly eccentric orbits. We will see below that the present self-consistent models produce even fewer highly eccentric orbits.

\citet{Elmegreen2016ApJ...830..115E} showed in a random walk model that stars experiencing random scatterings in a disc reach an equilibrium state distribution that is 1/r times an exponential when the scattering has an inward bias. \citet{Struck2019MNRAS.489.5919S} then proved in the context of equilibria of collisionless systems that a surface density profile of the same form is the zero entropy gradient solution to the cylindrically symmetric Poisson equation.
Because scattering processes prefer a maximum entropy, scattering nudges a disc towards a profile of 1/r times an exponential, which is very close to an exponential over a disc except for the innermost region where the factor of 1/r could make a noticeable difference.

Here we pursue the idea of stellar scattering by massive clumps, and for the first time use \textit{N}-body simulations to study 
its effect on the stellar density profile evolution. We want to know whether an exponential is still favoured as predicted by analytic derivation and what the differences are in comparison with the previous models.

\section{Model}
\label{sec:model}

In this section, we explain the procedures used to initialize and evolve the model galaxies. In order to focus on
the effect of gravitational scattering on profile evolution, many physics processes,
such as star formation, stellar feedback, and black hole feedback, are excluded, although some of them can play an important role on the profile evolution \citep[see][]{Struck2018ApJ...868L..15S}.

Three types of particles are used in our models to represent the populations of dark matter, star, and clumps, which work as massive scattering centres. After a proper initial condition gets specified for each galactic component, galaxies evolve under the control of the \textit{N}-body simulation code 
{\small GADGET}-2 designed by \citet{2005gadget}.

\subsection{Initial conditions}
\label{sec:ini cond} 

We build our initial galaxies mainly following the approach described by \citet{Springel2005MNRAS.361..776S}, with major modifications on the
surface density distribution of the disc. The initial profiles
of galaxy components are summarized in the following paragraphs.

The \citet{Hernquist1990} profile is used for the distribution of dark matter:
\begin{equation}
    \rho_{\text{dm}}(r) =\frac{M_{\text{dm}}}{2\pi} \frac{a}{r(r+a)^3},
	\label{eq:dark matter dist}
\end{equation}
where $M_{\text{dm}}$ is the mass of dark matter and $a$ is a constant.
Stars are initially placed to make a Gaussian disc. The surface density profile of stars is 
described by, 
\begin{equation}
    \Sigma_{\text{star}}(r) =\frac{M_{\text{star}}}{\pi h^2} ~ e^{-r^2/h^2},
	\label{eq:star dist}
\end{equation}
where the scale length $h$ controls the size of the stellar disc.
The reason of starting with a Gaussian disc instead of other nonexponential forms, e.g. a flat disc, is to avoid a sudden truncation of 
the density profile at the edge of the disc. The truncation may cause unrealistic edge instability and 
produce fast spread of outer stars into the vacant space, which hampers estimations of the rate of
profile evolution. 
The vertical mass distribution of stars is an isothermal profile. The initial three-dimensional stellar 
density profile is given by,
\begin{equation}
    \rho_{\text{star}}(r,z) =\frac{M_{\text{star}}}{2\pi z_0 h^2} ~ \text{sech}^2(\frac{z}{z_0 })~ e^{-r^2/h^2},
	\label{eq:star 3D dist}
\end{equation}
where $M_{\text{star}}$ is the mass of all the stars and $z_0$ is a scale height constant.
Clumps are set in the stellar disc using a inverse square root law distribution with a cutoff at the end
of the stellar disc. Mathematically, the density profile of clumps is written as,
\begin{equation}
    \Sigma_{\text{clump}} (r)=
   \begin{cases}
     \frac{3 M_{\text{clump}}}{4\pi b^{3/2}} ~ \frac{1}{\sqrt{r}}~, \quad   & 0< r <b\\
     0 ~,  & r > b 
   \end{cases}
	\label{eq:clump dist}
\end{equation}
where $M_{\text{clump}}$ is the mass of all the clumps, and the constant $b$ is determined by the radius of the stellar disc. We use this inverse square root profile to distribute clumps 
more or less uniformly in the disc with a small concentration at the disc centre.

The velocity profiles of galaxy components are chosen to make gravitational forces balanced with
centrifugal forces and the whole system close to an equilibrium. To do this, we adopt the procedure used by \citet{Springel2005MNRAS.361..776S}. We mention in particular that the assumption,
\begin{equation}
   \sigma_R^2 = \langle v_R^2\rangle = f_R \langle v_z^2 \rangle
	\label{eq:fr}
\end{equation}
 is taken for
the stellar disc, where $v_R$ and $v_z$ are stellar velocities in the radial direction and the vertical direction, and $\sigma_R$ is the radial velocity dispersion of stars. 

One limitation of the initial conditions described here is the modestly imperfect centripetal balance between gravitational forces and centrifugal forces.
As discussed later in the paper, ring waves induced from a degree of imbalance can disturb the analysis on orbital eccentricities and the radial bias of stellar scattering.
Studies on a control run, however, show that this limitation on the initial condition does not significantly affect our major results (see below).


We also employ the constant factor $f_R$ to control the Toomre $Q$ parameter of the initial disc, using the well-known relation,
\begin{equation}
   Q = \frac{\sigma_R \kappa}{3.36 G \Sigma},
	\label{eq:Toom Q}
\end{equation}
where $\kappa$ is the epicycle frequency. Like stars, clumps all have rotational velocities, which make them orbit around the galactic centre.

\subsection{Parameters in simulations}

\begin{figure}
	\includegraphics[width=\columnwidth]{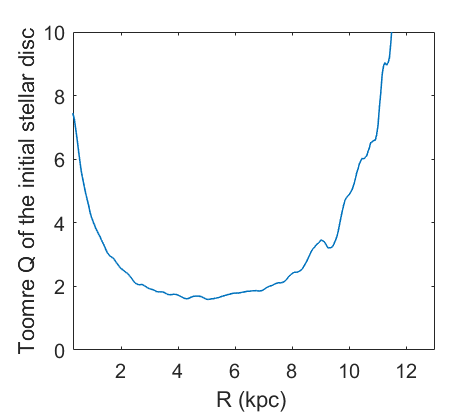}
    \caption{ Toomre $Q$ of the stellar disc at T = 0 in the fiducial run. The average value from $R$ = 3 kpc to $R$ = 8 kpc is $Q$ = 1.8.
    }
    \label{fig:initial q}
\end{figure}

Table~\ref{tab:para table} shows main parameters used in our fiducial run. The total mass of the fiducial model 
is in the range of a dwarf galaxy. We choose a dwarf galaxy, in part, as an analogue to a young galaxy. The scalelength $h$ in the equation \ref{eq:star dist} is chosen to be 5.4 kpc, which leads to an initial radius of about 15 kpc for the stellar disc given that the number of star particles is 1 million. The Gaussian profile makes more than 80\% of stars reside less than 7 kpc from the disc centre. The cutoff radius $b$ in the equation \ref{eq:clump dist} is set to 13 kpc to cover most of the stellar disc. The constant $a$ for halo Hernquist profile is 8.05 kpc. With the choices above, dark matter, stars, and clumps contribute to 77\% , 16\%, and 7\% of the mass within one disc scalelength. The contribution of each component to the initial rotation curve is discussed more below (cf. Figure \ref{fig: v_split}). Dark matter is dominant throughout the disc. 
The mass of an individual clump in the fiducial run is about $7.3 \times 10^{6} M_{\sun}$, which is close to the interstellar Jeans mass and the mass of the largest molecular clouds \citep{Elmegreen1987ApJ...320..182E,Williams1997ApJ...476..166W}. 
The mass of a star particle is about $1.5 \times 10^{3} M_{\sun}$. Since this value is significantly greater than the masses of real stars, 
the gravitational field around a stellar particle is stronger than a real star. This indicates
the cross section of stellar collisions is greater than the real value. To counteract this effect,
we choose a relatively large gravitational softening length
for stars to ensure a collisionless stellar model. The radial dispersion factor $f_R$ in the equation \ref{eq:fr} is set to 1, which makes the initial radial dispersion the same as the vertical one. 
Figure~\ref{fig:initial q} shows Toomre $Q$ at T = 0 as a function of galactic radii in the fiducial run. The average value from $R$ = 3 kpc to $R$ = 8 kpc is $Q$ = 1.8. This relatively large $Q$ makes the stellar disc dynamically hot, reducing the effect of resonant scattering at corotation.

Parameters used in a control run, which is designed to confirm that a profile evolution in the fiducial run is mainly caused by massive clumps, are identical to the parameters in the table~\ref{tab:para table} , except for a change on the number of clumps from 
200 to 1 000 000. This modification makes the mass of individual clump the same as a star particle so that
stellar scattering due to clumps is not pronounced. The mass of all the clumps stays unchanged to keep 
the gravitational potential from clumps controlled.

The sets of parameters in other runs are based on our fiducial run. When studying the effect of the mass
of an individual clump on the timescale of the profile evolution, we alter only the number 
of clumps and keep the total mass of all the clumps unchanged, so that the individual mass varies and the total
gravitational potential almost remains constant. 
When studying the effect of Toomre $Q$ of the stellar disc, we alter only the parameter $f_R$ which appears in the equation~\ref{eq:fr}. $f_R$  changes the radial dispersion, which is related to Toomre $Q$ according to the equation \ref{eq:Toom Q}.

The time unit used in all the simulations is 49 Myr.

\begin{table}
	\centering
	\caption{Main parameters in the fiducial run}
	\label{tab:para table}
	\begin{tabular}{lc} 
		\hline
		Parameters & Values\\
		\hline
		Total mass $M_{\text{tot}}$ & $2.9 \times 10^{10} M_{\sun}$\\
		$M_{\text{dm}}/M_{\text{tot}}$ & 0.90 \\
		$M_{\text{star}}/M_{\text{tot}}$ & 0.05 \\
		$M_{\text{clump}}/M_{\text{tot}}$ & 0.05 \\
		Number of dark matter particles $N_{\text{dm}}$ & 970 000\\
		Number of star particles $N_{\text{star}}$ & 1 000 000\\
		Number of clumps $N_{\text{clump}}$ & 200\\
		Dark halo Hernquist constant $a$ & 8.05 kpc\\ 
		Initial scalelength of the stellar disc $h$ & 5.4 kpc\\
		Initial scale height of the stellar disc $z_0$ &  0.45 kpc\\
		Cutoff radius of the disc of clumps $b$ & 13 kpc\\
		Radial dispersion factor $f_R$ &  1\\
		Toomre $Q$ of the initial stellar disc $Q_{i} $ & 1.8\\ 
		Gravitational softening for dark matter $\epsilon_{\text{dm}}$ & 0.2 kpc\\
		Gravitational softening for stars $\epsilon_{\text{star}}$ & 0.1 kpc\\
		Gravitational softening for clumps $\epsilon_{\text{clump}}$ & 0.1 kpc\\	
		\hline
	\end{tabular}
\end{table}

\section{Results}

\subsection{Formation of an exponential disc}

\begin{figure}
	\includegraphics[width=\columnwidth]{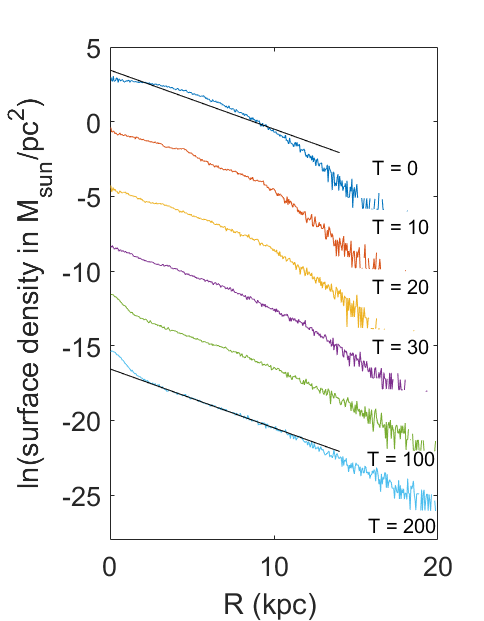}
    \caption{Evolution of the stellar surface density profile in the fiducial run. 
    Each successive profile, except for T = 0, is offset downwards by 4 to give clear views.
    Two black lines, which have the same slope, are added to make the change in profiles easier
    to see. The time unit is 49 Myr.
    }
    \label{fig:sdp change}
\end{figure}

As shown in Figure~\ref{fig:sdp change},
the surface density profile evolves from an Gaussian towards an exponential.  
Two black lines with the same slope and length are shown to make clear the profile difference between T = 0 and T = 200.
At T = 0, the black line does not fit the initial Gaussian disc.
The line intersects the profile at $R = 2.5$ kpc and $R = 10$ kpc, so   
the stellar density at T = 0 is above the line in the range between 2.5 kpc and
10 kpc. 
At T = 200, the black line fits very well from 2 kpc to 10 kpc. At this time, the profile beyond 10 kpc can be fit by another line with a steeper slope, so we claim that the profile is a Type II exponential, which is composed of an inner exponential and an steeper outer exponential. The profile within $R = 2$ kpc forms a small cusp, where the density is 
above the black line. This cusp can be seen at earlier times. As shown at T = 30 and T = 100, the central cusp emerges after the profile of the inner disc becomes exponential.
From T = 0 to T = 200, the net change in stellar density can be roughly interpreted in
the way that some stars in the range between 2.5 kpc and
10 kpc relocate at the innermost part of the disc. Comparisons among profiles at different times show that the profile evolution occurs at a 
fast rate before T = 30 and settles down afterwards. The difference in profile between T = 100 and T = 200 is little. 


\begin{figure*}
	\includegraphics[width=\textwidth]{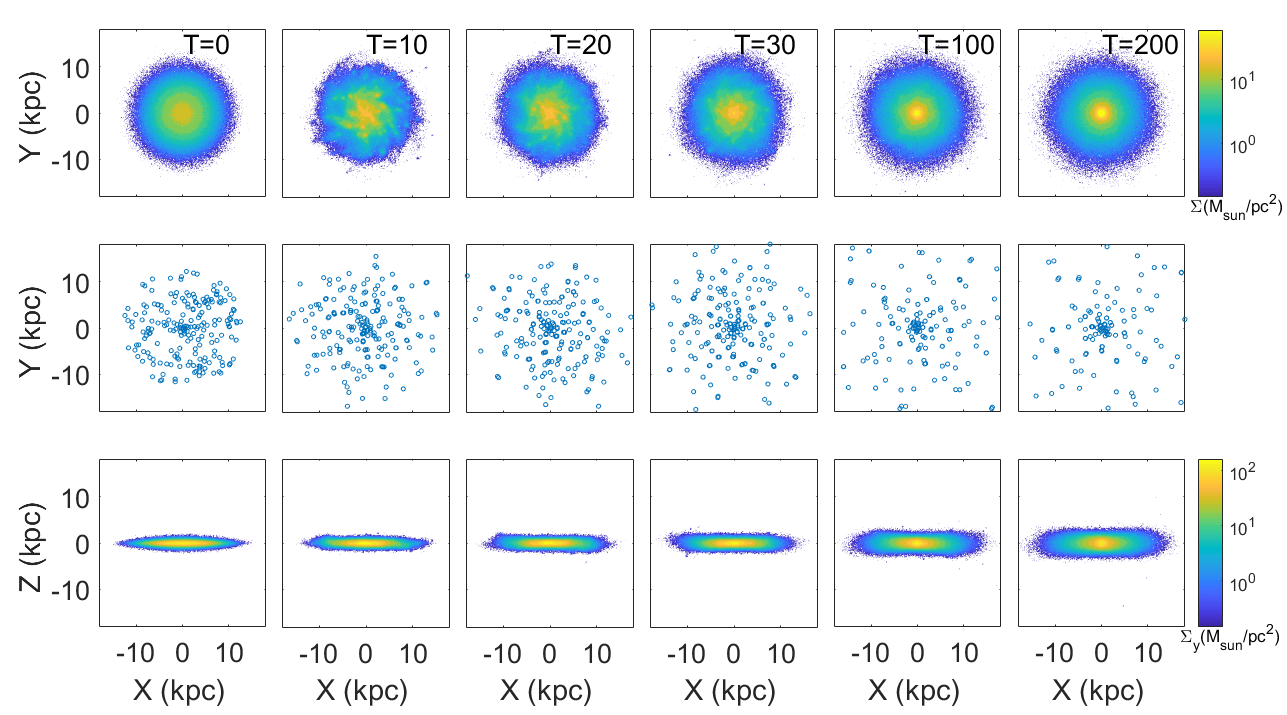}
    \caption{Evolution of the stellar disc and clumps in the fiducial run. The first row shows the 
    top view of the stellar disc. The second row shows the top view of 200 clumps. The third row 
    shows the side view of the stellar disc. Colors used in the first row and the third row reflect stellar mass density viewed face-on and edge-on respectively. The time unit is 49 Myr.
    }
    \label{fig:all snaps}
\end{figure*}

The formation of the exponential is accompanied by other interesting processes. Figure~\ref{fig:all snaps} shows the changing appearance of the stellar disc and clumps during the profile
evolution. On some top view panels (e.g., at T = 10) of the stellar disc,
one can see little dotted overdense regions, which result from 
stars clustering around massive clumps owing to their strong gravitational attraction. The clustering enhances the individual clump mass and can promote stellar scattering.
At T = 10 and T = 20, flocculent spiral arms are excited in the stellar disc and mild ring waves propagate outwards from the centre. The appearance of arms can be explained by disc density inhomogeneities introduced by orbiting clumps \citep{D'Onghia2013ApJ...766...34D}. The ring waves, which also reveal themselves as bumps on surface density curves in Figure \ref{fig:sdp change} (e.g., around $R = 4$ kpc at T = 10), are caused by a degree of non-equilibrium in initial conditions. As we will see later in the discussion on the control run, the ring waves should not affect most of the surface density evolution.
The arms 
and waves gradually die out at later times in the fiducial run. Panels in the second row 
reveal the redistribution of clumps. As the stellar density profile alters, some clumps near the edge of the disc move further away, and
inner clumps get slightly concentrated in the disc centre. These
motions of clumps are discussed more below (cf. Figure \ref{fig: v_split}). From T = 0 to T = 40, the contribution to the rotation velocity due to clump gravity increases in the region $R$ < 8 kpc due to central concentration, while the contribution declines at $R$ > 11 kpc as a result of clump diffusion at the edge. The concentration of clumps was also observed in the simulation done by \citet{Bournaud2007}, where the increasing compactness at the centre led to the formation of a bulge.
Panels in the third row illustrate a disc thickening.
More discussion on the disc thickening is in Section~\ref{sec: thickening}.

\begin{figure*}
	\includegraphics[width=\textwidth]{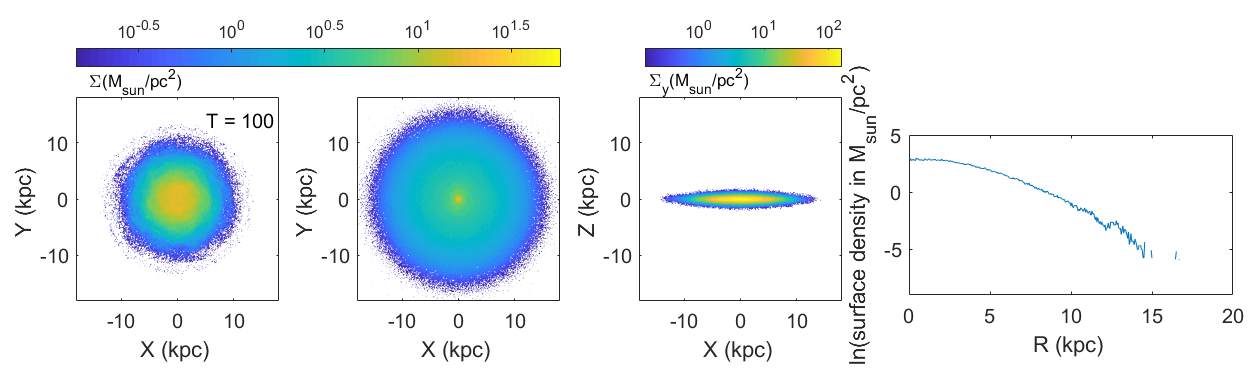}
    \caption{The galactic disc and the stellar surface profile in the control run at T = 100. Panels from left to right are the 
    top view of the stellar disc, the top view of clumps, the side view of the stellar disc, and 
    the stellar surface profile, respectively. Colors used in the face-on views and the edge-on view reflect the mass density of the viewed component. The time unit is 49 Myr.
    }
    \label{fig: control snaps }
\end{figure*}

The formation of the exponential profile results from stellar scattering off from massive clumps. 
Figure~\ref{fig: control snaps } shows the results of our control run, where the number of clumps
increases to one million and the mass of all the clumps remains unvaried.
The individual clump mass is identical to a single star, so the scattering from clumps is minimal in this run. At T = 100, the stellar surface stays a Gaussian distribution in spite of the presence of  a ring at $R = 13$ kpc. The distinct results of the control run show that the profile 
evolution in the fiducial model is mainly caused by massive clumps rather than ring waves.
The change in the vertical profile of the stellar disc is not observed in the control run.

\begin{figure}
	\includegraphics[width=\columnwidth]{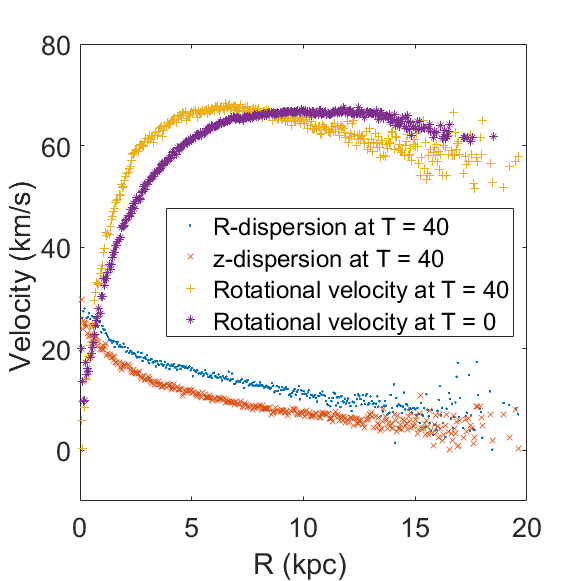}
    \caption{Velocity profiles of the fiducial run. Blue dots, orange crosses, and yellow plus signs
    represent the radial velocity dispersion, the vertical velocity dispersion, and the rotational 
    velocities, respectively, at T = 40. The purple asterisks show the rotational velocities at T = 0.
    All the velocities are azimuthal averages.
    }
    \label{fig: v vs r}
\end{figure}

\begin{figure}
	\includegraphics[width=\columnwidth]{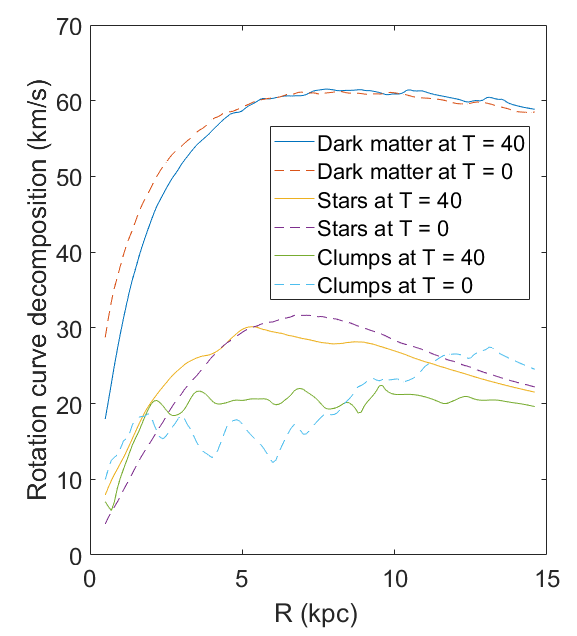}
    \caption{ Azimuthally averaged rotation curves for different components in the fiducial run. Solid lines represent curves at T = 40 and dashed lines represent T = 0. The unevenness seen in the clump curves results from large variance in gravitational potentials at different azimuthal angles, as the number of clumps in the run is relatively low.
    }
    \label{fig: v_split}
\end{figure}

Figure~\ref{fig: v vs r} shows the velocity profiles of the fiducial model.
From T = 0 to T = 40, rotational velocities slightly rise in the inner region of the disc. The rotation curve decomposition in Figure~\ref{fig: v_split} shows that this rise comes from the inward relocation of disc stars and clumps. To clarify this change in rotation curve, 
consider a thin uniform ring of stars
at $R = 5$ kpc moving to the location of $R = 1$ kpc. When the ring is at $R = 5$ kpc, the net gravitational force acting on stars in the annulus between 1 kpc and 5 kpc points outward. When the ring moves to $R = 1$ kpc, the gravitational force from the ring in this region becomes inward. In other words, the shrinkage of the stellar ring makes the total centripetal force in the annulus increase. Therefore the rotational velocities in the annulus have to rise to maintain a centripetal balance. 
In Figure~\ref{fig: v vs r}, the radial velocity dispersion and the vertical velocity dispersion both decline with increasing  
galactic radius. This decrease looks like an exponential profile and the profile agrees with the expectation
made by \citet{Martinsson2013} and \citet{Struck2019MNRAS.489.5919S}. More discussion on the velocity dispersion is in Section~\ref{sec:comp analytic}.

\begin{figure}
	\includegraphics[width=\columnwidth]{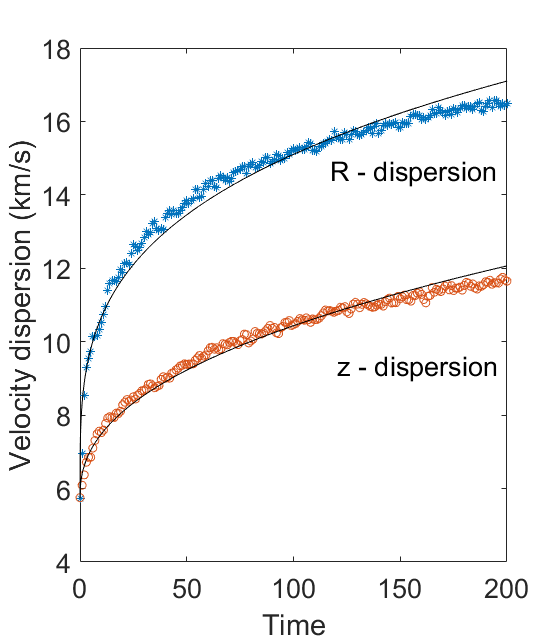}
    \caption{ Time evolution of velocity dispersions in the fiducial model. Blue asterisks and
    orange circles represent the radial dispersion and the vertical dispersion respectively.
    The plotted values at each time are dispersions averaged in the annulus between 5 kpc and
    10 kpc. Black curves are the best fitting curves assuming a power-law model that the velocity dispersion is proportional to $t^\alpha$. The fitting gives $\alpha = 0.28 \pm 0.01$ for the $R$-dispersion and  $\alpha = 0.43 \pm 0.01$ for the $z$-dispersion.
    }
    \label{fig: disp vs t}
\end{figure}

Figure~\ref{fig: disp vs t} shows time evolution of velocity dispersion in the fiducial model.
The radial dispersion starts around 6 km/s and reaches approximately 16.5 km/s at T = 200.  
The vertical dispersion starts around 6 km/s and ends up around 11.7 km/s at T = 200.  
When compared with the rotational velocity of 65 km/s, the dispersions  at T = 200 seem reasonable.
The increase of the dispersion in both directions is rapid at the beginning and gradually slows down. The shape of the two curves resemble a function of $t^{\alpha}$ with $\alpha = 0.28 \pm 0.01$ for the $R$-dispersion and  $\alpha = 0.43 \pm 0.01$ for the $z$-dispersion. This result is consistent with the analytic studies on the influence of
giant clouds on stellar velocity dispersion \citep{Lacey1984,Spitzer1953, Villumsen1983ApJ...274..632V}. \citet{Aumer2017MNRAS.470.3685A} showed age-velocity dispersion relation (AVR)
of the solar neighbourhood observed by the Geneva-Copenhagen Survey on their Figure 11. Although our galaxy model is not designed to simulate the Milky Way, both the trend of our velocity dispersion evolution and the ratio between radial dispersion and vertical dispersion agree with the observed AVR.
The increase in velocity dispersion is a reflection of disc heating.
The rise in $z$-dispersion is connected to disc thickening.
More discussion on the disc thickening is in Section~\ref{sec: thickening}.

\subsection{Factors that affect profile evolution timescale}

The fiducial model illustrates that stellar scattering due to massive clumps can be an effective way to generate an
exponential disc. The exponential profile starts to appear about 1.5 Gyr after the start of the run. In this subsection, we study what factors can affect the rate of profile evolution. We find the 
evolution timescale is influenced by the mass of an individual clump and the disc stability.

\begin{figure}
	\includegraphics[width=\columnwidth]{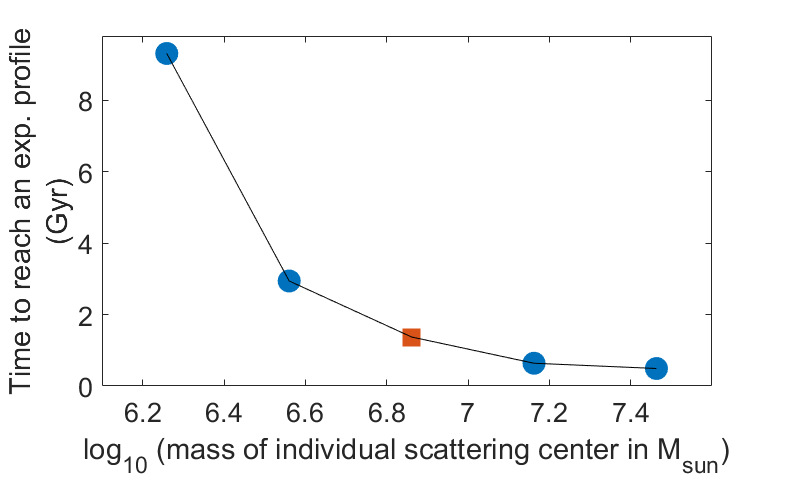}
    \caption{ Time to reach an exponential profile as a function of individual clump mass.
    Each marker in the graph represents a run, with the orange square representing the 
    fiducial run. All the five runs use the same set of parameters, except for the number of clumps. Since the total mass of all the clumps is kept unchanged, variations in the number of clumps
alter the individual clump mass. 
    }
    \label{fig:mass time}
\end{figure}

High mass of an individual clump can accelerate the formation of an exponential.
Figure~\ref{fig:mass time} shows the times to reach an exponential in five runs, with the red square representing the fiducial run.
More specifically, times in the figure are the moments when an inner exponential fully emerges for the first time in a run right before the formation of a central cusp. 
These five runs take the same set of initial conditions and parameters, except the number of clumps.
Since the total mass of all the clumps is kept unchanged, variations in the number of clumps
alter the individual clump mass. As one can see, the time to reach an exponential is more sensitive
to the individual mass in the mass range less than the fiducial value. When the mass deceases by a factor of 4 with respect to the fiducial run, the time goes from 1.5 Gyr to 9.5 Gyr. When the mass increases by a factor of 4, the time becomes 0.5 Myr, reducing to one third of the fiducial value. Further increase in individual mass does not noticeably shorten the time.
The effect of weak individual scattering in a low-clump-mass case can dominate 
the increase in the number of scattering centres, making the overall profile evolution slow.
High-mass clumps can induce the clustering of stars around them as seen in the fiducial run, which makes the scattering centres even more massive. Lastly, a relatively small number of clumps in a high-mass 
run tends to break the axial symmetry of the spatial distribution of the clumps. 
A small number of clumps covers the space of azimuthal positions with large gaps in azimuthal angles. This makes the gravitational field coming from all clumps have some non-axisymmetric features.
For example, an azimuthally uniform distribution of six clumps at the same galactic radius creates an $m = 6$ pattern in the clump potential. A non-uniform distribution can generate even lower orders of harmonic patterns.  
Since we know that non-axisymmetries in spiral arms and $m = 2$ bars are capable of scattering stars and producing an exponential profile, 
the non-axisymmetry in the gravitational field of clumps may play the same role and therefore expedite the density profile change.

\begin{figure}
	\includegraphics[width=\columnwidth]{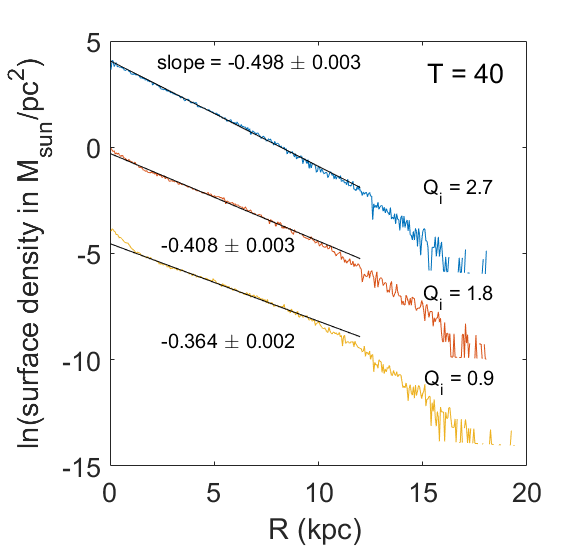}
    \caption{ Surface density at T = 40 for three runs starting with different radial 
    velocity dispersion profiles. The Toomre $Q$ of the initial stellar disc for the blue
    run, the orange run, and the yellow run are 2.7, 1.8, and 0.9 respectively. The orange run 
    is the fidicual run. Black lines are linear regression lines that fit each density 
    profile in the range from 2 kpc to 10 kpc. The slopes of these three lines are shown in the figure.
    Each successive profile, except the blue (top) one, is offset downwards by 4 to give clear views.
    }
    \label{fig:q}
\end{figure}

Discs with low stability also tend to evolve faster. The Toomre $Q$ parameter is used as a measure 
of disc stability. Three simulations with different initial radial velocity dispersions are 
conducted and their surface profiles at T = 40 are shown in Figure~\ref{fig:q}. The best-fitting exponential profiles over the range from 2 kpc to 10 kpc are shown.
The slopes of these three profiles from top to bottom are $-0.498\pm 0.003$, $-0.408\pm0.003$, and $-0.364\pm0.002$, showing that the exponentials that three runs will
finish with have slightly different scale lengths and that the scale length corresponding to a higher initial
Toomre $Q_i$ is slightly shorter. The difference in slope makes it difficult to determine the time to reach
an exponential, but we can still compare the rate of profile evolution among these runs by looking at the size of a central cusp. As shown before in Figure~\ref{fig:sdp change}, a cusps starts
to develop after an inner exponential forms.
The well pronounced central cusp in the yellow ($Q_i = 0.9$) run indicates that the density profile of this run has evolved to a more advanced stage than the other two runs. Discs with lower initial  Toomre $Q$ are able to evolve slightly faster. 
A lower $Q$ makes the disc more reactive to perturbations, so the secondary effects of clumping around the clumps and spiral wavelets are larger at a lower $Q$, which can expedite the profile evolution. The strong clumping can also scatter stars more intensely,  throwing them further out, thereby explaining the shallower slope in a low-$Q$ run. The trial of $Q_i = 0.9$ is initially unstable. It develops manifest arms and the disc gets heated up rapidly. However, in this case, the migration of stars still occurs and the density profile still evolves to an exponential. 

\section{Discussion}

\subsection{Changes in stellar orbits}

\subsubsection{Eccentricity}

The density profile evolution in a galaxy by nature is a collective effect of changes in billions of stellar orbits. Preliminary analyses on stellar orbits show that most of the abrupt variations on orbits in our simulations are caused by close encounters with one or more clumps, although some may result from interactions with spiral wavelets. Here we discuss some properties of stellar orbits. More detailed analysis will be done in the future.

\begin{figure*}
	\includegraphics[width=\textwidth]{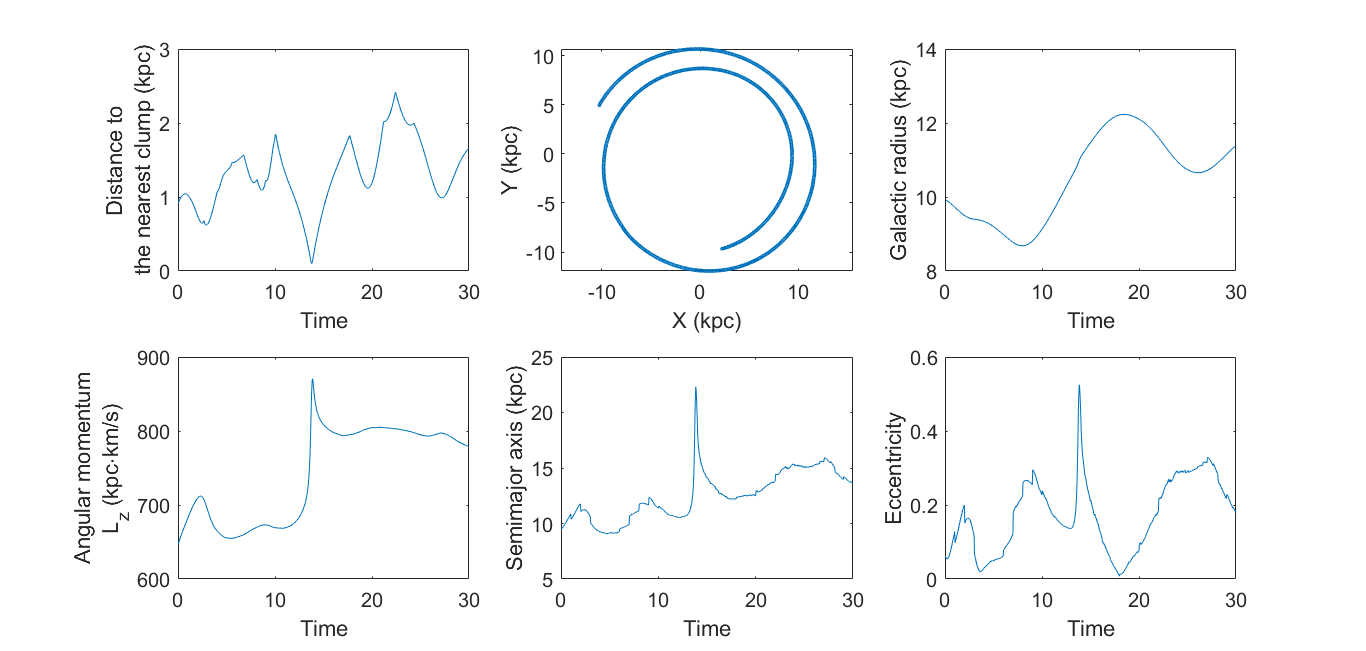}
    \caption{Orbital parameters of a star in the fiducial run from T = 0 to T = 30. The top left panel gives the distance from this star to the nearest clump at a given time. The top middle panel shows the projection of the star's orbit on the midplane of the galaxy. The star starts at (2.24, -9.66), moves counterclockwise, and reaches (-10.3, 4.95) at T = 30. The top right panel shows the distance from the star to the galactic centre. 
    The three bottom panels from left to right show the $z$-component of the specific angular momentum, the semi-major axis, and the eccentricity of the star, respectively. The  semi-major axis and the eccentricity at each time point are  estimated by the specific angular momentum and the specific energy at that time assuming a Keplerian orbit fit as a first order approximation.  }
    \label{fig:orbit}
\end{figure*}

Figure~\ref{fig:orbit} shows orbital parameters of a star in the fiducial run.
The top left panel gives the distance from this star to the nearest clump as a function of time.
As indicated by the big dip on this panel,
the star experiences an encounter with clumps around T = 13 . The three bottom panels from left to right show the time dependence of the $z$-component of the specific angular momentum, the semi-major axis, and the eccentricity of the star, respectively. The  semi-major axis and the eccentricity at each time point are  estimated by the specific angular momentum and the specific energy at that time assuming a Keplerian orbit to the first order. 
When the star encounters one or more clumps around T = 13, sharp changes in the stellar velocity produce dramatic temporary variations on calculated angular momentum, semi-major axis, and eccentricity as seen in the figure. Neglecting the spike on the eccentricity panel which occurs at the encounter time, the net eccentricity of the orbit decreases moderately after the scattering. The angular momentum, however, changes significantly after the encounter at T = 13. This type of clump scatterings, which can shift stars radially without an increase in eccentricities, is common in the fiducial run. \citet{Sellwood2002} mentioned,
in the appendix of their paper, that a star cloud scattering produces little change in stellar random motion energy if a star passes a cloud whose azimuthal speed relative to the star is close to 0. This theory is a possible explanation for the scattering we see here. 

\begin{figure*}
	\includegraphics[width=\textwidth]{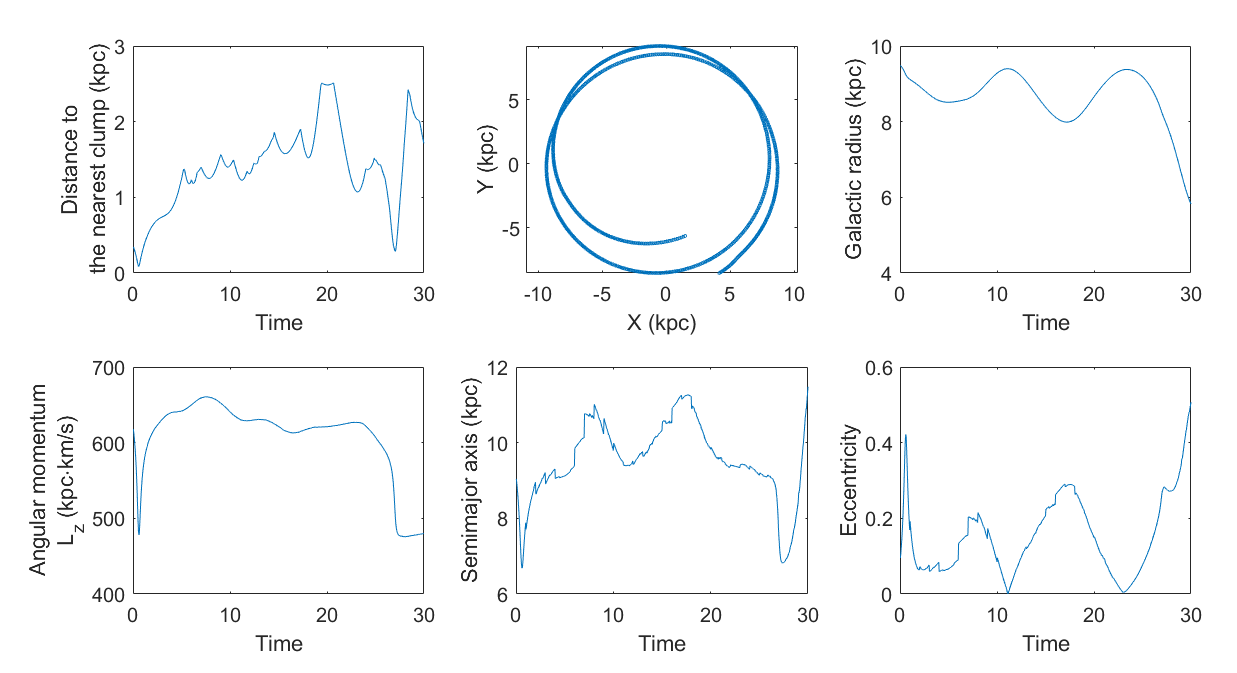}
    \caption{Orbital parameters of another star in the fiducial run from T = 0 to T = 30. Quantities plotted in this figure are the same as Figure~\ref{fig:orbit}. This star starts at (4.15, -8.53), moves counterclockwise, and reaches (1.47, -5.64) at T = 30.  }
    \label{fig:orbit2}
\end{figure*}

Close encounters can generate a more eccentric orbit sometime. Figure~\ref{fig:orbit2} shows orbital parameters of another star in the fiducial run. The big dip on the top left panel indicates that an encounter happens around T = 27. As shown in the angular momentum panel and the eccentricity panel, this encounter pulls the star inward and makes the orbit more eccentric. This type of scatterings contributes to a rise in the average orbital eccentricity, as we will see in the next figure. 

\begin{figure*}
	\includegraphics[width=\textwidth]{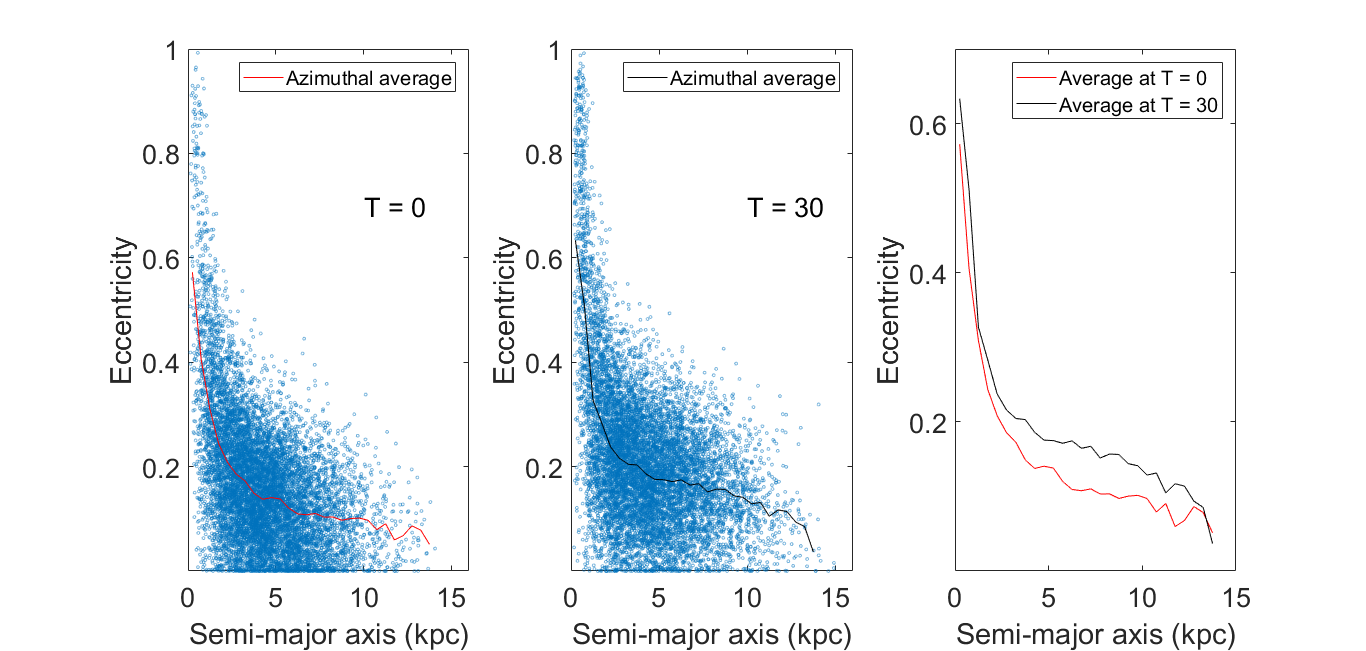}
    \caption{Orbital eccentricities of stars
as a function of semi-major axis in our fiducial model. The left and the middle panel shows the distribution of eccentricities at T = 0 and T = 30 respectively. Each blue dot represents a star. 
Red lines and black lines show
eccentricities averaged over semi-major axis bins. The panel on the right show the two average
lines again, using a smaller scale on the y-axis, to make an easy comparison. }
    \label{fig:eccentricity}
\end{figure*}

Figure~\ref{fig:eccentricity} shows the distribution of stellar orbital eccentricities
and semi-major axes at T = 0 and T = 30 in our fiducial model. Each blue dot represents a star. 
Red lines and black lines show
eccentricities averaged over semi-major axis bins. The panel on the right shows the two average
lines again, using a smaller scale on the y-axis, to make an easy comparison. In general, stars in our 
model have eccentricities less than 0.4. Stars with very high eccentricities are located only in the inner
disc. 
From T = 0 to T = 30, the eccentricities of stars go up slightly in the annulus from 3 to 13 kpc, which is in accord with the increase in the velocity dispersions shown in Figure~\ref{fig: disp vs t}.


\begin{figure*}
	\includegraphics[width=\textwidth]{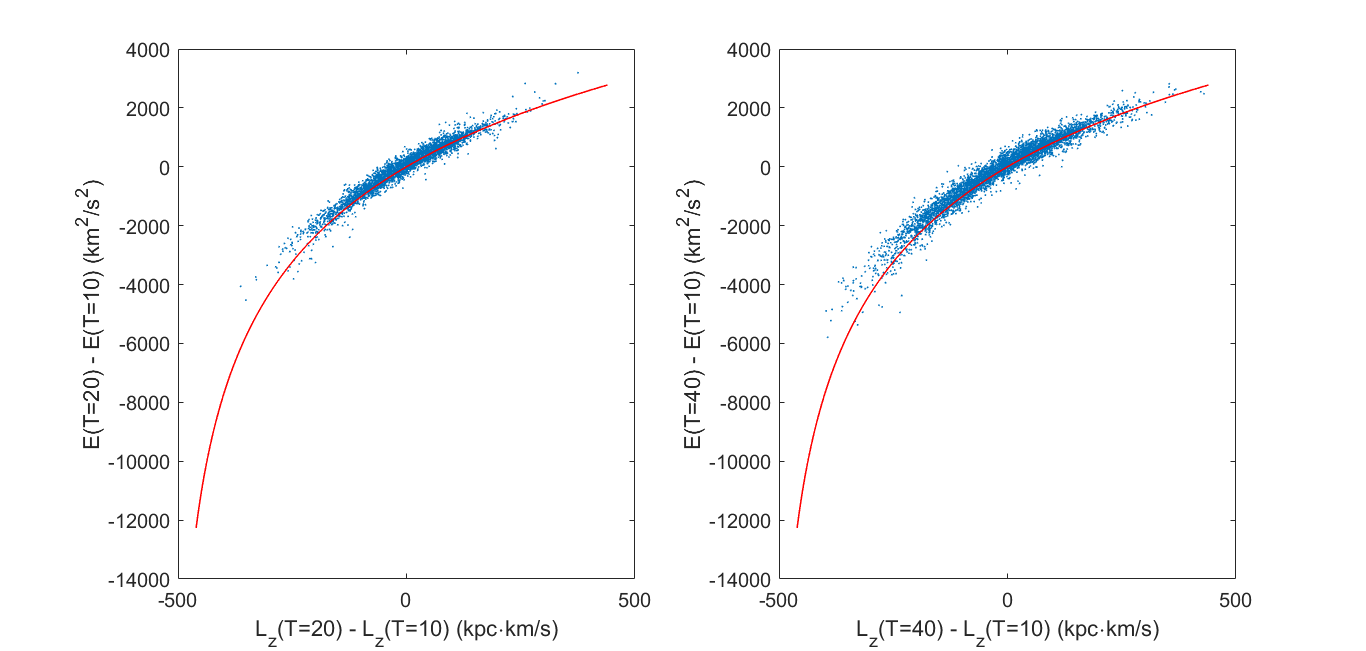}
    \caption{ Changes in the specific mechanical energy and the z-component of specific angular momentum for 6534 stars sampled with galactic radii between 7.5 kpc and 7.6 kpc at T = 10 in the fiducial run. The top panel and the bottom panel show the changes from T = 10 to T = 20 and from T = 10 to T = 40 respectively. In both panels, each blue dot represents one sampled star, and the red solid lines show how the energy and the angular momentum would change if a sampled star were in a circular orbit at T = 10 and maintained its circularity on an orbit of different size at a later time.
     }
    \label{fig:LE}
\end{figure*}

\subsubsection{Energy and angular momentum}

Figure~\ref{fig:LE} shows changes in the specific mechanical energy and the z-component of specific angular momentum for 6534 stars sampled with galactic radii between 7.5 kpc and 7.6 kpc at time T = 10 in the fiducial run.
In both panels, each blue dot represents one sampled star, and the red solid lines show how the energy and the angular momentum would change if a sampled star were in a circular orbit at T = 10 and maintained its circularity on an orbit of different size at a later time. Changes in specific angular momentum reflect shifts of orbits in the radial direction, or orbit size. At 7.5 kpc, an increase of 200 kpc$\cdot$km/s in $L_z$ corresponds to an outwards shift of the guiding center radius by 3 kpc, and an increase of 400 kpc$\cdot$km/s gives a shift of almost 6 kpc. Since the red lines in both panel match well with the distribution of blue dots, we say that radial shifts of most orbits are not accompanied by large increases in random orbital energy. When taking a closer look at the figure, the vertical span of the distribution at a given abscissa, however, does increase slightly from the left panel to the right panel, indicating that there is a small increase in random energy on average. This is consistent with the moderate change in average eccentricity. 
In the figure, the right end of the distribution converges on the red line better than the left end, indicating that stars moving outwards are more likely to conserve orbital circularity than stars moving in.


The comparison between the two panels in the figure also shows that shifts of stellar orbits correlate with time. A longer time allows stars to move further away from their original paths.

\subsection{Disc thickening}
\label{sec: thickening}
As seen in Figure~\ref{fig:all snaps}, disc thickening is a  characteristic of the fiducial run. Here we discuss some details on vertical profiles of the fiducial run.

\begin{figure}
	\includegraphics[width=\columnwidth]{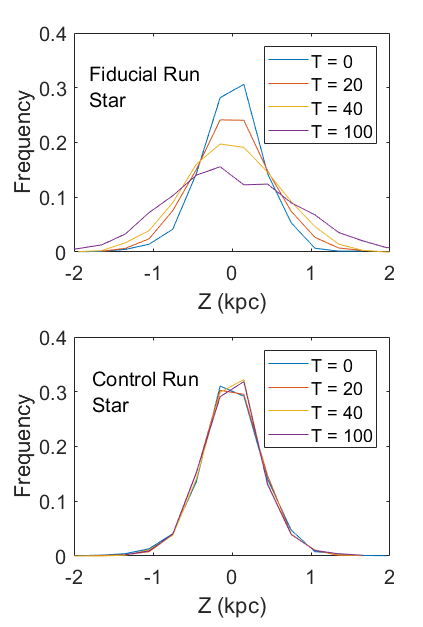}
    \caption{ Normalized vertical distributions of stars at different times in the fiducial run (top panel) and the control run (bottom panel). Blue, orange, yellow, and purple lines (from highest peak to lowest) represent distributions at T = 0, 20, 40, and 100, respectively. Stars are sampled in a region with galactic radii between 3 kpc and 7 kpc. The z-coordinates of stars are binned using a bin width of 0.3 kpc. 
     }
    \label{fig:z_distribute}
\end{figure}

As shown in Table~\ref{tab:para table}, the stellar disc starts with a scale height $z_0$ of 0.45 kpc, assuming a hyperbolic secant squared (sech$^2(z/z_0$)) profile.  Figure~\ref{fig:z_distribute} shows how the scale height within galactic radii between 3 kpc and 7 kpc evolves with time in the fiducial run (top panel) and in the control run (bottom panel). Clearly in the fiducial run, stars spread away from the mid-plane of the disc, with the scale height gradually increasing and reaching about 1.1 kpc at T = 100. In the control run, however, the scale height hardly changes with time. To figure out reasons behind disc thickening in the fiducial run, it is necessary to look at the vertical profile of clumps. As close encounters with clumps are responsible for most changes in stellar orbits, the vertical distribution of clumps can significantly affect vertical motions of stars. 

\begin{figure}
	\includegraphics[width=\columnwidth]{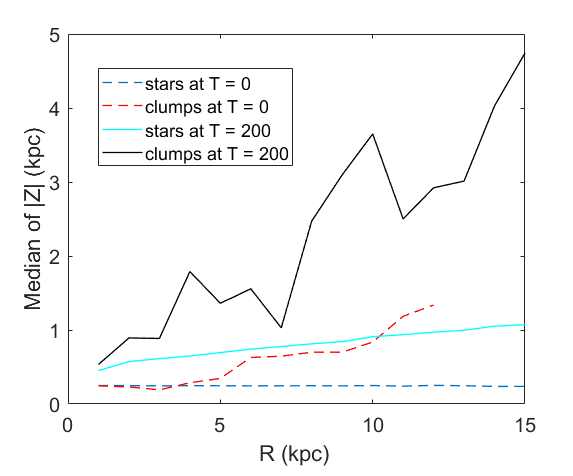}
    \caption{ Half-mass heights of stars and clumps as a function of galactic radii at T = 0 and T = 200. The blue dashed line and the red dashed line represent median values of |Z| at different radii at T = 0 for stars and clumps respectively. The cyan line and the black line are median values of |Z| at T = 200.
     }
    \label{fig:H_vs_r}
\end{figure}

Figure~\ref{fig:H_vs_r} shows half-mass heights of stars and clumps as a function of galactic radii at T = 0 and T = 200. When the secant squared distribution is assumed in $z$-direction, a half-mass height is about 55\% of the scale height $z_0$. Since $z_0 = 0.45 $ kpc at T = 0, the half-mass height for the initial stellar disc is 0.25 kpc at all radii. The half-mass heights of clumps at T = 0 are almost identical to those of stars within 5 kpc, but the clump disc flares in the outer region. This initial vertical profile of clumps can be viewed as mimicking the distribution of gas in an accretion process. 
Since the half-mass heights of clumps are initially no less than stars in most regions of the disc, there are many clumps moving with a vertical amplitude that is greater than most stars. Scatterings from these clumps can pump stars to higher altitudes. At T = 200, half-mass heights of stars and clumps are much higher than the initial values. The reason why clumps in the outer disc get high vertical displacement may be that the density of the stellar disc there is too low to effectively constrain the vertical motion of the clumps.

The half-mass heights of stars at this time show a near-linear dependence on galactic radii. From $ R = 1 $ kpc to $ R = 15 $ kpc, the height increases by a factor of 2 as shown in Figure~\ref{fig:H_vs_r}. This aligns very well with the results by \citet{Narayan2002A&A...390L..35N}. The reason we do not get a constant height, as seen in \citet{Bournaud2009ApJ...707L...1B} and \citet{Wilson2019ApJ...882....5W}, may be that our galaxy is dominated by dark matter at all radii. When using the equation
\begin{equation}
    z_0=\frac{\sigma_z^2}{\pi G \Sigma_{\text{star}}}
	\label{eq:H}
\end{equation}
derived by \citet{vander1981A&A....95..105V} to calculate the scale height of stars at T = 200, the scale height obtained rises up linearly in the inner disc, but goes exponentially beyond 6 kpc. 
In the inner disc, the vertical velocity dispersion decreases nearly exponentially with radius (see Figure~\ref{fig: v vs r} for reference, the dispersion profile at T = 200 has a shape similar to that at T = 40). This decrease balances with the exponential stellar surface density, making the calculated $z_0$ have a very low rate of increase within 6 kpc. In the outer disc, the decrease of the dispersion is much slower (see Figure~\ref{fig: v vs r}), and the decrease of the surface density is faster (see Figure~\ref{fig:sdp change}). Therefore, the calculated $z_0$ rises up exponentially in the outer disc. A more accurate way to compute $z_0$ is to replace the stellar surface density $\Sigma_{\text{star}}$ in the equation \ref{eq:H} with the total surface density from all the galactic components. When dark matter and clumps were included, the surface density would decrease like an inverse power law in the outer disc, causing $z_0$ to increase much less rapidly in the outer part and fit the curve in the figure better.


\subsection{Comparison with previous simulations}

Results in our simulations confirm most of those of the test particle models of \citet{Struck2017}. In their 
models, a Type II exponential forms out of a flat initial disc due to stellar scattering, and
a noticeable disc thickening accompanying the profile evolution is seen. 
Moreover, the developed velocity profiles and the evolution of velocity dispersion 
in their simulations are similar to ours. This consistency verifies the validity of 
non-self-gravitating models used to investigate the effect of stellar scattering.
There are also some differences. The first one comes when comparing the distribution of orbital eccentricity of stars. Most stars in their model have eccentricities greater than 0.4, which is much higher
than the eccentricities shown in our Figure~\ref{fig:eccentricity}. On one hand, as they mentioned in their paper, the very high eccentricities may be an artefact due to the unreal initial conditions in their model. On the other hand, high eccentricities in their model indicate that the nature of stellar scattering in their run may be different than ours (e.g., with different smoothing sizes). 
The second difference is the time to reach an exponential.
Their fudicial model takes a longer time to obtain a smooth exponential. However, since parameters and initial conditions used are so different, the evolution time is not directly comparable.

\citet{Bournaud2007} showed, using a sticky-particle code, that a gas-rich flat disc
evolves to an exponential disc in 1 Gyr with the rapid formation of clumpy structures along
the way. While stellar scattering due to massive clumps plays an important role in their simulations regarding the formation of the exponential, two other factors also significantly
contribute to this process. The first one is that the initial disc in their fiducial run
is gravitationally unstable. This instability leads to violent phase mixing and
produces intense grand-design spirals. Phase mixing may assist in making 
exponentials. The other factor is that their initial density profile has a sharp edge at 6 kpc. The sudden truncation of the density profile is not realistic and causes fast 
spread of the outer disc, which dominates the surface profile evolution in the region beyond
6 kpc. In our models, we circumvent these two factors and try focusing on the effect of stellar
scattering only.

\subsection{Comparison with analytic theories}
\label{sec:comp analytic}

\begin{figure}
	\includegraphics[width=\columnwidth]{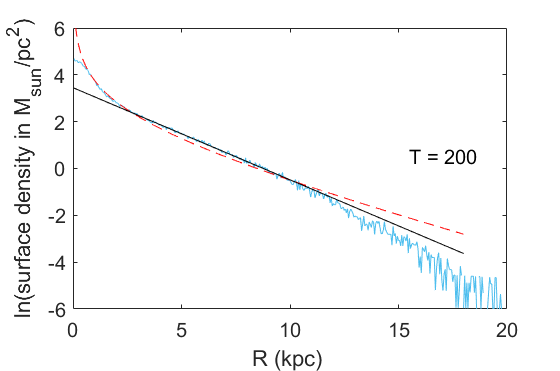}
    \caption{ Surface density profile at T = 200 in the fiducial run. The red dashed line shows the best fitting curve with a form of 1/r times an exponential. The galactic radii within in 12.5 kpc are used for this fitting. The black line shows a linear fitting, which represents a purely exponential profile.
     }
    \label{fig:better_fit}
\end{figure}

\citet{Struck2019MNRAS.489.5919S} did analytic work, showing that the zero entropy gradient solution to the Poisson equation 
has the form of 1/r times an exponential for: surface density, radial velocity dispersion, and the square of vertical velocity dispersion in the case of self-gravitating discs. 
Stellar scattering pushes the disc towards this  solution regardless of the initial
shape of the disc. Our models are in line with most of their results. The appearance of the central
cusp in our models makes the final surface profile resemble a function of 1/r at the innermost
disc, and the radial decrease in the rest of the disc resembles an exponential, so the overall density profile fits 1/r times an exponential well. 
Figure~\ref{fig:better_fit} shows the best fitting curve with a form of 1/r times an exponential at T = 200 in the fiducial run. From 3 kpc to 11 kpc, the difference between the red dashed curve and the black line is little. In the region $R < 3$ kpc, 1/r times an exponential matches the density profile much better. Therefore, we claim that 1/r times an exponential is preferred for the eventual equilibrium state of the disc. We have to say that the central cusp may not appear in the surface profile of a real galaxy, because the central region has large mean free paths and other physical processes excluded by our model may dominate there. 
Velocity dispersions in Figure~\ref{fig: v vs r} show a near-exponential decrease in the inner disc, but the rate of the decrease slows down in the outer disc. The digression from an exponential decrease in the outer disc may indicate a degree of incomplete scattering in our model.

\begin{figure}
	\includegraphics[width=\columnwidth]{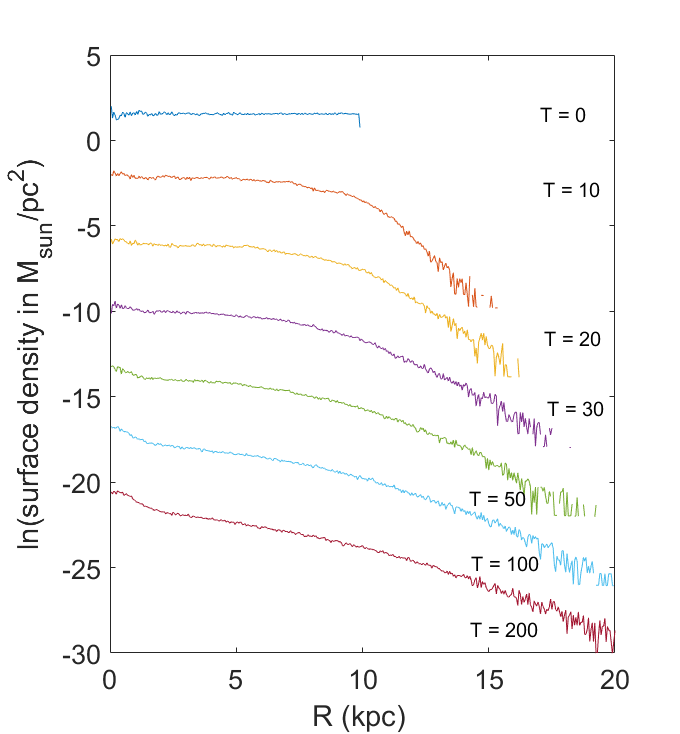}
    \caption{ A run starting with a flat density profile reaches a near-exponential profile with a central cusp.
    }
    \label{fig:flat initial}
\end{figure}

\begin{figure}
	\includegraphics[width=\columnwidth]{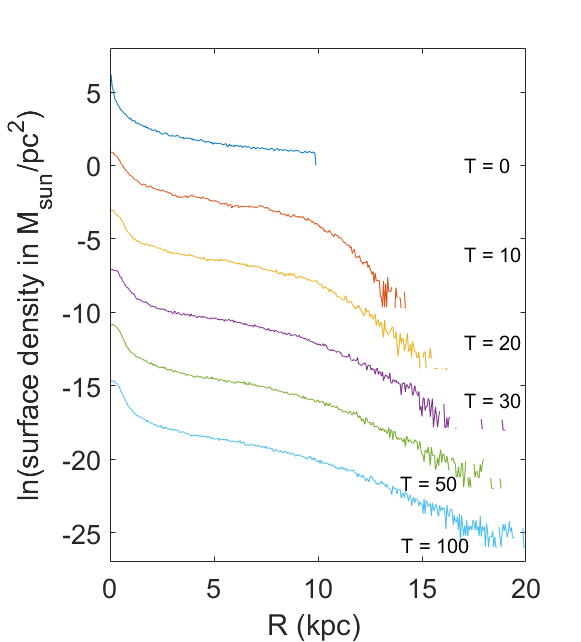}
    \caption{ Another run starting with an 1/r density profile. This model reaches an near-exponential profile with a central cusp as well.
    }
    \label{fig:mestel initial}
\end{figure}

We did a few special runs with non-Gaussian initial profiles to test the dependence of profile
evolution on initial density distributions. 
Figure~\ref{fig:flat initial} shows the profile evolution of a flat disc.
At T = 200, the disc exhibits a near-exponential profile with a central cusp. Figure~\ref{fig:mestel initial} shows a profile evolution starting with an 1/r density profile. This model reaches a near-exponential profile with a central cusp as well. These results seem to support that 1/r times an exponential is favoured by scattering under various initial profiles. 

\begin{figure}
	\includegraphics[width=\columnwidth]{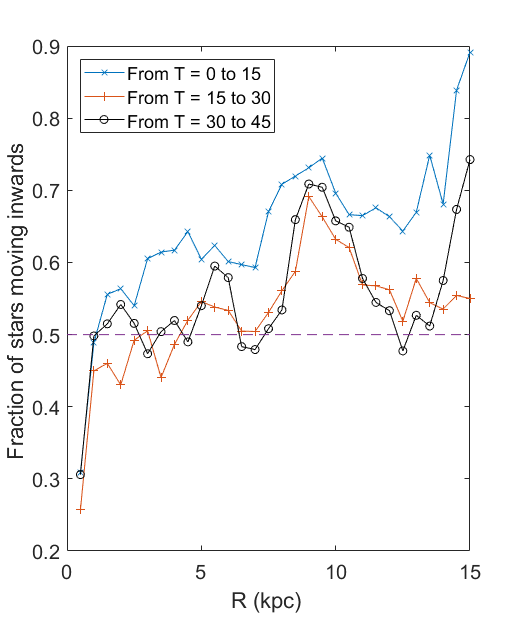}
    \caption{ Fraction of stars moving inwards in the fiducial run. Stars in 30 thin annuli with different galactic radii are tracked for a period of 15 time units. A star's final location is compared with its initial location to determine whether the star moves inwards or not. The x-axis represents galactic radii of sampled annuli, and the y-value of each mark shows the fraction of stars in that annulus moving inwards during the period. Blue
    crosses, orange plus signs, and black open circles represent time periods from 0 to 15, from 
    15 to 30, and from 30 to 45, respectively. The purple dash line shows the fraction equal to 0.5, where the number of stars moving inwards equals stars moving outwards.
    }
    \label{fig:bias}
\end{figure}

\citet{Elmegreen2016ApJ...830..115E} showed that when stars in a disc experience multiple two-dimensional scatterings and the relocation of stars due to the scatterings prefers inwards
directions than outwards directions by a fixed degree, the density distribution of stars will converge to
1/r times an exponential, which is a stable equilibrium under a two-dimensional random walk model. The scale length of the exponential is determined by the strength of the inwards
bias. A stronger bias generates a steeper exponential.
They also showed that if the strength of the bias is dependent of scattering locations in the disc, the density profile can evolve to a double exponential. More specifically, if the inwards bias in the inner disc is less intense than the outer disc, 
a Type II exponential will form. If the inner disc is more intense, a Type III 
exponential will occur. Stellar scatterings in our fiducial model have directional biases. To measure the strength of biases and examine its dependence on galactic radius , we track stars in 30 thin annuli for a period of 750 Myr and use the excess fraction of stars moving inwards versus
outwards as an index for bias strength. As shown in Figure~\ref{fig:bias}, the fractions of 
stars moving inwards exhibit both time dependence and radial dependence. 

During the period from T = 0 to 15, the fractions in almost all the radii are greater than 0.5, indicating an inwards bias across the entire disc. The fraction at 0.5 kpc is less than 0.5 because it is
almost impossible for these innermost stars to move further in. As time goes on, the inwards bias of scattering diminishes. During the periods from 15 to 30 and from 30 to 45, the disc within 7 kpc has no apparent bias. Since the profile evolution towards
an exponential occurs fast from T = 0 to 15, the formation of the exponential in our model can result from the inwards scattering bias seen in this first period. As shown by the wave-like shape of the three curves in the figure, the fractions of stars moving inwards oscillate as functions of radii. This may be due to an  axisymmetric disc vibration mode in radial direction coming from the imperfect initial centripetal balance. This secondary radial oscillation superposed on the relatively global inward bias can produce the curves. 

\section{Conclusions}

Results from self-gravitating simulations show that stellar scattering due to massive clumps can lead to the formation of an exponential disc.
Scattering increases stellar velocity dispersions, and the time evolution of radial and vertical dispersions resemble a function of $t^{\alpha}$ with $\alpha = 0.28 \pm 0.01$ for the radial dispersion and  $\alpha = 0.43 \pm 0.02$ for the vertical dispersion, which agrees with analytic studies on the influence of giant gas clouds on stars. The disc thickens with time because of scattering.
This is unlike resonant radial migration which does not lead to disc thickening \citep{Minchev2012A&A...548A.127M}.
 The  scale height of the evolved stellar disc increases linearly by about a factor of 2 from the centre to the edge. Eccentricities of stellar orbits increase modestly during density profile evolution. Close encounters between stars and clumps can make stellar orbits more eccentric as strong impulsive scattering events seen in \citet{Struck2017}, but many leave the stellar eccentricities with little change, which is similar to the cloud scattering with a low passing speed as discussed in the appendix of \citet{Sellwood2002}. Resonant radial migration is unlikely to be responsible, as \citet{Vera-Ciro2014ApJ} found that scattering at corotation resonance (churning) does not change the profile or other structural parameters of the disc. We are currently investigating these processes in more detail. Other processes, like multiple scattering events, may play important roles in the profile evolution as well.


The rate of profile evolution due to scattering correlates with local gravitational stability of the stellar disc and is sensitive to the masses of individual scattering centres. Low stability and high scattering mass can generate exponential forms quickly. Since Toomre $Q$
used to measure disc stability is related to stellar surface density, a change in disc mass or disc size will affect the evolution timescale. Other factors, such as the size of clumps and the mass of the dark matter halo, may have an influence on the timescale as well. Our fiducial model reaches a Type II exponential in about 1.5 Gyr, but this time can vary as properties of the galaxy change.

The profile evolution due to scattering can make a near-exponential disc under various initial density distributions. The appearance of a central cusp in the final profile makes the equilibrium distribution resemble 
a function of 1/r times an exponential, like that derived by analytic work assuming a minimum entropy gradient of the stellar system \citep{Struck2019MNRAS.489.5919S, Elmegreen2016ApJ...830..115E} or assuming a maximum entropy in specific angular momentum distribution \citep{herpich2017MNRAS.467.5022H}. \citet{Marr2020Galax...8...12M} also showed that an equilibrium galactic disc is a microcanonical ensemble with maximum entropy and that the equilibrium disc density profile is a lognormal distribution, which can fit with a near-exponential profile well.
The central cusp may not be realized in galaxies with relatively large mean free paths in the central regions, since these analyses assume scattering on all scales. Similarly in the outer parts of discs, quite long-lived Type II and III profiles may result from incomplete scattering evolution or driving by another component, such as a bar. Nonetheless, with these caveats, the convergence of these different approaches, and the present work, suggest this profile form is robust and universal.

\section*{Acknowledgements}

We thank Dr. Stephen Pardy for suggestions on simulation codes and Dr. Fred Adams for comments made on our results.

\section*{Data availability}

The data used in this article will be shared on request to the corresponding author.




\bibliographystyle{mnras}
\bibliography{reference} 

\begin{thebibliography}{}
\makeatletter
\relax
\def\mn@urlcharsother{\let\do\@makeother \do\$\do\&\do\#\do\^\do\_\do\%\do\~}
\def\mn@doi{\begingroup\mn@urlcharsother \@ifnextchar [ {\mn@doi@}
  {\mn@doi@[]}}
\def\mn@doi@[#1]#2{\def\@tempa{#1}\ifx\@tempa\@empty \href
  {http://dx.doi.org/#2} {doi:#2}\else \href {http://dx.doi.org/#2} {#1}\fi
  \endgroup}
\def\mn@eprint#1#2{\mn@eprint@#1:#2::\@nil}
\def\mn@eprint@arXiv#1{\href {http://arxiv.org/abs/#1} {{\tt arXiv:#1}}}
\def\mn@eprint@dblp#1{\href {http://dblp.uni-trier.de/rec/bibtex/#1.xml}
  {dblp:#1}}
\def\mn@eprint@#1:#2:#3:#4\@nil{\def\@tempa {#1}\def\@tempb {#2}\def\@tempc
  {#3}\ifx \@tempc \@empty \let \@tempc \@tempb \let \@tempb \@tempa \fi \ifx
  \@tempb \@empty \def\@tempb {arXiv}\fi \@ifundefined
  {mn@eprint@\@tempb}{\@tempb:\@tempc}{\expandafter \expandafter \csname
  mn@eprint@\@tempb\endcsname \expandafter{\@tempc}}}

\bibitem[\protect\citeauthoryear{{Aumer}, {Binney}  \& {Sch{\"o}nrich}}{{Aumer}
  et~al.}{2017}]{Aumer2017MNRAS.470.3685A}
{Aumer} M.,  {Binney} J.,   {Sch{\"o}nrich} R.,  2017, \mn@doi [\mnras]
  {10.1093/mnras/stx1483}, \href
  {https://ui.adsabs.harvard.edu/abs/2017MNRAS.470.3685A} {470, 3685}

\bibitem[\protect\citeauthoryear{{Boroson}}{{Boroson}}{1981}]{boroson1981}
{Boroson} T.,  1981, \mn@doi [\apjs] {10.1086/190742}, \href
  {https://ui.adsabs.harvard.edu/abs/1981ApJS...46..177B} {46, 177}

\bibitem[\protect\citeauthoryear{{Bournaud}, {Elmegreen}  \&
  {Elmegreen}}{{Bournaud} et~al.}{2007}]{Bournaud2007}
{Bournaud} F.,  {Elmegreen} B.~G.,   {Elmegreen} D.~M.,  2007, \mn@doi [\apj]
  {10.1086/522077}, \href
  {https://ui.adsabs.harvard.edu/abs/2007ApJ...670..237B} {670, 237}

\bibitem[\protect\citeauthoryear{{Bournaud}, {Elmegreen}  \&
  {Martig}}{{Bournaud} et~al.}{2009}]{Bournaud2009ApJ...707L...1B}
{Bournaud} F.,  {Elmegreen} B.~G.,   {Martig} M.,  2009, \mn@doi [\apjl]
  {10.1088/0004-637X/707/1/L1}, \href
  {https://ui.adsabs.harvard.edu/abs/2009ApJ...707L...1B} {707, L1}

\bibitem[\protect\citeauthoryear{{Bowler}, {Dunlop}, {McLure}  \&
  {McLeod}}{{Bowler} et~al.}{2017}]{Bowler2017}
{Bowler} R.~A.~A.,  {Dunlop} J.~S.,  {McLure} R.~J.,   {McLeod} D.~J.,  2017,
  \mn@doi [\mnras] {10.1093/mnras/stw3296}, \href
  {https://ui.adsabs.harvard.edu/abs/2017MNRAS.466.3612B} {466, 3612}

\bibitem[\protect\citeauthoryear{{D'Onghia}, {Vogelsberger}  \&
  {Hernquist}}{{D'Onghia} et~al.}{2013}]{D'Onghia2013ApJ...766...34D}
{D'Onghia} E.,  {Vogelsberger} M.,   {Hernquist} L.,  2013, \mn@doi [\apj]
  {10.1088/0004-637X/766/1/34}, \href
  {https://ui.adsabs.harvard.edu/abs/2013ApJ...766...34D} {766, 34}

\bibitem[\protect\citeauthoryear{{Daniel} \& {Wyse}}{{Daniel} \&
  {Wyse}}{2018}]{Daniel2018MNRAS.476.1561D}
{Daniel} K.~J.,  {Wyse} R. F.~G.,  2018, \mn@doi [\mnras]
  {10.1093/mnras/sty199}, \href
  {https://ui.adsabs.harvard.edu/abs/2018MNRAS.476.1561D} {476, 1561}

\bibitem[\protect\citeauthoryear{{Daniel}, {Schaffner}, {McCluskey}, {Fiedler
  Kawaguchi}  \& {Loebman}}{{Daniel} et~al.}{2019}]{Daniel2019ApJ...882..111D}
{Daniel} K.~J.,  {Schaffner} D.~A.,  {McCluskey} F.,  {Fiedler Kawaguchi} C.,
  {Loebman} S.,  2019, \mn@doi [\apj] {10.3847/1538-4357/ab341a}, \href
  {https://ui.adsabs.harvard.edu/abs/2019ApJ...882..111D} {882, 111}

\bibitem[\protect\citeauthoryear{{Elmegreen} \& {Elmegreen}}{{Elmegreen} \&
  {Elmegreen}}{1987}]{Elmegreen1987ApJ...320..182E}
{Elmegreen} B.~G.,  {Elmegreen} D.~M.,  1987, \mn@doi [\apj] {10.1086/165534},
  \href {https://ui.adsabs.harvard.edu/abs/1987ApJ...320..182E} {320, 182}

\bibitem[\protect\citeauthoryear{{Elmegreen} \& {Elmegreen}}{{Elmegreen} \&
  {Elmegreen}}{2005}]{EE2005}
{Elmegreen} B.~G.,  {Elmegreen} D.~M.,  2005, \mn@doi [\apj] {10.1086/430514},
  \href {https://ui.adsabs.harvard.edu/abs/2005ApJ...627..632E} {627, 632}

\bibitem[\protect\citeauthoryear{{Elmegreen} \& {Struck}}{{Elmegreen} \&
  {Struck}}{2013}]{Elmegreen2013ApJ...775L..35E}
{Elmegreen} B.~G.,  {Struck} C.,  2013, \mn@doi [\apjl]
  {10.1088/2041-8205/775/2/L35}, \href
  {https://ui.adsabs.harvard.edu/abs/2013ApJ...775L..35E} {775, L35}

\bibitem[\protect\citeauthoryear{{Elmegreen} \& {Struck}}{{Elmegreen} \&
  {Struck}}{2016}]{Elmegreen2016ApJ...830..115E}
{Elmegreen} B.~G.,  {Struck} C.,  2016, \mn@doi [\apj]
  {10.3847/0004-637X/830/2/115}, \href
  {https://ui.adsabs.harvard.edu/abs/2016ApJ...830..115E} {830, 115}

\bibitem[\protect\citeauthoryear{{Elmegreen}, {Elmegreen}, {Vollbach}, {Foster}
   \& {Ferguson}}{{Elmegreen} et~al.}{2005}]{Elmegreen2005ApJ...634..101E}
{Elmegreen} B.~G.,  {Elmegreen} D.~M.,  {Vollbach} D.~R.,  {Foster} E.~R.,
  {Ferguson} T.~E.,  2005, \mn@doi [\apj] {10.1086/496952}, \href
  {https://ui.adsabs.harvard.edu/abs/2005ApJ...634..101E} {634, 101}

\bibitem[\protect\citeauthoryear{{F{\"o}rster Schreiber} et~al.,}{{F{\"o}rster
  Schreiber} et~al.}{2006}]{Forster2006ApJ...645.1062F}
{F{\"o}rster Schreiber} N.~M.,  et~al., 2006, \mn@doi [\apj] {10.1086/504403},
  \href {https://ui.adsabs.harvard.edu/abs/2006ApJ...645.1062F} {645, 1062}

\bibitem[\protect\citeauthoryear{{Freeman}}{{Freeman}}{1970}]{Freeman1970ApJ...160..811F}
{Freeman} K.~C.,  1970, \mn@doi [\apj] {10.1086/150474}, \href
  {https://ui.adsabs.harvard.edu/abs/1970ApJ...160..811F} {160, 811}

\bibitem[\protect\citeauthoryear{{Hernquist}}{{Hernquist}}{1990}]{Hernquist1990}
{Hernquist} L.,  1990, \mn@doi [\apj] {10.1086/168845}, \href
  {https://ui.adsabs.harvard.edu/abs/1990ApJ...356..359H} {356, 359}

\bibitem[\protect\citeauthoryear{{Herpich}, {Tremaine}  \& {Rix}}{{Herpich}
  et~al.}{2017}]{herpich2017MNRAS.467.5022H}
{Herpich} J.,  {Tremaine} S.,   {Rix} H.-W.,  2017, \mn@doi [\mnras]
  {10.1093/mnras/stx352}, \href
  {https://ui.adsabs.harvard.edu/abs/2017MNRAS.467.5022H} {467, 5022}

\bibitem[\protect\citeauthoryear{{Herrmann}, {Hunter}  \&
  {Elmegreen}}{{Herrmann} et~al.}{2013}]{Herrmann2013AJ....146..104H}
{Herrmann} K.~A.,  {Hunter} D.~A.,   {Elmegreen} B.~G.,  2013, \mn@doi [\aj]
  {10.1088/0004-6256/146/5/104}, \href
  {https://ui.adsabs.harvard.edu/abs/2013AJ....146..104H} {146, 104}

\bibitem[\protect\citeauthoryear{{Hohl}}{{Hohl}}{1971}]{hohl1971}
{Hohl} F.,  1971, \mn@doi [\apj] {10.1086/151091}, \href
  {https://ui.adsabs.harvard.edu/abs/1971ApJ...168..343H} {168, 343}

\bibitem[\protect\citeauthoryear{{Ichikawa}, {Wakamatsu}  \&
  {Okamura}}{{Ichikawa} et~al.}{1986}]{ichikawa1986}
{Ichikawa} S.~I.,  {Wakamatsu} K.~I.,   {Okamura} S.,  1986, \mn@doi [\apjs]
  {10.1086/191094}, \href
  {https://ui.adsabs.harvard.edu/abs/1986ApJS...60..475I} {60, 475}

\bibitem[\protect\citeauthoryear{{Lacey}}{{Lacey}}{1984}]{Lacey1984}
{Lacey} C.~G.,  1984, \mn@doi [\mnras] {10.1093/mnras/208.4.687}, \href
  {https://ui.adsabs.harvard.edu/abs/1984MNRAS.208..687L} {208, 687}

\bibitem[\protect\citeauthoryear{{Lin} \& {Pringle}}{{Lin} \&
  {Pringle}}{1987}]{Lin1987}
{Lin} D.~N.~C.,  {Pringle} J.~E.,  1987, \mn@doi [\apjl] {10.1086/184981},
  \href {https://ui.adsabs.harvard.edu/abs/1987ApJ...320L..87L} {320, L87}

\bibitem[\protect\citeauthoryear{{Marr}}{{Marr}}{2020}]{Marr2020Galax...8...12M}
{Marr} J.~H.,  2020, \mn@doi [Galaxies] {10.3390/galaxies8010012}, \href
  {https://ui.adsabs.harvard.edu/abs/2020Galax...8...12M} {8, 12}

\bibitem[\protect\citeauthoryear{{Martinsson}, {Verheijen}, {Westfall},
  {Bershady}, {Schechtman-Rook}, {Andersen}  \& {Swaters}}{{Martinsson}
  et~al.}{2013}]{Martinsson2013}
{Martinsson} T. P.~K.,  {Verheijen} M. A.~W.,  {Westfall} K.~B.,  {Bershady}
  M.~A.,  {Schechtman-Rook} A.,  {Andersen} D.~R.,   {Swaters} R.~A.,  2013,
  \mn@doi [\aap] {10.1051/0004-6361/201220515}, \href
  {https://ui.adsabs.harvard.edu/abs/2013A&A...557A.130M} {557, A130}

\bibitem[\protect\citeauthoryear{{Mestel}}{{Mestel}}{1963}]{Mestel1963}
{Mestel} L.,  1963, \mn@doi [\mnras] {10.1093/mnras/126.6.553}, \href
  {https://ui.adsabs.harvard.edu/abs/1963MNRAS.126..553M} {126, 553}

\bibitem[\protect\citeauthoryear{{Minchev} \& {Famaey}}{{Minchev} \&
  {Famaey}}{2010}]{Minchew2010}
{Minchev} I.,  {Famaey} B.,  2010, \mn@doi [\apj]
  {10.1088/0004-637X/722/1/112}, \href
  {https://ui.adsabs.harvard.edu/abs/2010ApJ...722..112M} {722, 112}

\bibitem[\protect\citeauthoryear{{Minchev}, {Famaey}, {Quillen}, {Dehnen},
  {Martig}  \& {Siebert}}{{Minchev} et~al.}{2012}]{Minchev2012A&A...548A.127M}
{Minchev} I.,  {Famaey} B.,  {Quillen} A.~C.,  {Dehnen} W.,  {Martig} M.,
  {Siebert} A.,  2012, \mn@doi [\aap] {10.1051/0004-6361/201219714}, \href
  {https://ui.adsabs.harvard.edu/abs/2012A&A...548A.127M} {548, A127}

\bibitem[\protect\citeauthoryear{{Narayan} \& {Jog}}{{Narayan} \&
  {Jog}}{2002}]{Narayan2002A&A...390L..35N}
{Narayan} C.~A.,  {Jog} C.~J.,  2002, \mn@doi [\aap]
  {10.1051/0004-6361:20020961}, \href
  {https://ui.adsabs.harvard.edu/abs/2002A&A...390L..35N} {390, L35}

\bibitem[\protect\citeauthoryear{{Patterson} \& {Thuan}}{{Patterson} \&
  {Thuan}}{1996}]{Patterson1996}
{Patterson} R.~J.,  {Thuan} T.~X.,  1996, \mn@doi [\apjs] {10.1086/192357},
  \href {https://ui.adsabs.harvard.edu/abs/1996ApJS..107..103P} {107, 103}

\bibitem[\protect\citeauthoryear{{Pe{\~n}arrubia}, {McConnachie}  \&
  {Babul}}{{Pe{\~n}arrubia} et~al.}{2006}]{Pe2006ApJ...650L..33P}
{Pe{\~n}arrubia} J.,  {McConnachie} A.,   {Babul} A.,  2006, \mn@doi [\apjl]
  {10.1086/508656}, \href
  {https://ui.adsabs.harvard.edu/abs/2006ApJ...650L..33P} {650, L33}

\bibitem[\protect\citeauthoryear{{Pohlen} \& {Trujillo}}{{Pohlen} \&
  {Trujillo}}{2006}]{Pohlen2006}
{Pohlen} M.,  {Trujillo} I.,  2006, \mn@doi [\aap]
  {10.1051/0004-6361:20064883}, \href
  {https://ui.adsabs.harvard.edu/abs/2006A&A...454..759P} {454, 759}

\bibitem[\protect\citeauthoryear{{Ro{\v{s}}kar}, {Debattista}, {Stinson},
  {Quinn}, {Kaufmann}  \& {Wadsley}}{{Ro{\v{s}}kar}
  et~al.}{2008}]{Roskar2008ApJ...675L..65R}
{Ro{\v{s}}kar} R.,  {Debattista} V.~P.,  {Stinson} G.~S.,  {Quinn} T.~R.,
  {Kaufmann} T.,   {Wadsley} J.,  2008, \mn@doi [\apjl] {10.1086/586734}, \href
  {https://ui.adsabs.harvard.edu/abs/2008ApJ...675L..65R} {675, L65}

\bibitem[\protect\citeauthoryear{{S{\'a}nchez-Bl{\'a}zquez}, {Courty}, {Gibson}
   \& {Brook}}{{S{\'a}nchez-Bl{\'a}zquez}
  et~al.}{2009}]{sanchez2009MNRAS.398..591S}
{S{\'a}nchez-Bl{\'a}zquez} P.,  {Courty} S.,  {Gibson} B.~K.,   {Brook} C.~B.,
  2009, \mn@doi [\mnras] {10.1111/j.1365-2966.2009.15133.x}, \href
  {https://ui.adsabs.harvard.edu/abs/2009MNRAS.398..591S} {398, 591}

\bibitem[\protect\citeauthoryear{{Sellwood} \& {Binney}}{{Sellwood} \&
  {Binney}}{2002}]{Sellwood2002}
{Sellwood} J.~A.,  {Binney} J.~J.,  2002, \mn@doi [\mnras]
  {10.1046/j.1365-8711.2002.05806.x}, \href
  {https://ui.adsabs.harvard.edu/abs/2002MNRAS.336..785S} {336, 785}

\bibitem[\protect\citeauthoryear{{Spitzer} \& {Schwarzschild}}{{Spitzer} \&
  {Schwarzschild}}{1953}]{Spitzer1953}
{Spitzer} Lyman J.,  {Schwarzschild} M.,  1953, \mn@doi [\apj]
  {10.1086/145730}, \href
  {https://ui.adsabs.harvard.edu/abs/1953ApJ...118..106S} {118, 106}

\bibitem[\protect\citeauthoryear{{Springel}}{{Springel}}{2005}]{2005gadget}
{Springel} V.,  2005, \mn@doi [\mnras] {10.1111/j.1365-2966.2005.09655.x},
  \href {https://ui.adsabs.harvard.edu/abs/2005MNRAS.364.1105S} {364, 1105}

\bibitem[\protect\citeauthoryear{{Springel}, {Di Matteo}  \&
  {Hernquist}}{{Springel} et~al.}{2005}]{Springel2005MNRAS.361..776S}
{Springel} V.,  {Di Matteo} T.,   {Hernquist} L.,  2005, \mn@doi [\mnras]
  {10.1111/j.1365-2966.2005.09238.x}, \href
  {https://ui.adsabs.harvard.edu/abs/2005MNRAS.361..776S} {361, 776}

\bibitem[\protect\citeauthoryear{{Struck} \& {Elmegreen}}{{Struck} \&
  {Elmegreen}}{2017}]{Struck2017}
{Struck} C.,  {Elmegreen} B.~G.,  2017, \mn@doi [\mnras]
  {10.1093/mnras/stx918}, \href
  {https://ui.adsabs.harvard.edu/abs/2017MNRAS.469.1157S} {469, 1157}

\bibitem[\protect\citeauthoryear{{Struck} \& {Elmegreen}}{{Struck} \&
  {Elmegreen}}{2018}]{Struck2018ApJ...868L..15S}
{Struck} C.,  {Elmegreen} B.~G.,  2018, \mn@doi [\apjl]
  {10.3847/2041-8213/aaedb4}, \href
  {https://ui.adsabs.harvard.edu/abs/2018ApJ...868L..15S} {868, L15}

\bibitem[\protect\citeauthoryear{{Struck} \& {Elmegreen}}{{Struck} \&
  {Elmegreen}}{2019}]{Struck2019MNRAS.489.5919S}
{Struck} C.,  {Elmegreen} B.~G.,  2019, \mn@doi [\mnras]
  {10.1093/mnras/stz2555}, \href
  {https://ui.adsabs.harvard.edu/abs/2019MNRAS.489.5919S} {489, 5919}

\bibitem[\protect\citeauthoryear{{Vera-Ciro}, {D'Onghia}, {Navarro}  \&
  {Abadi}}{{Vera-Ciro} et~al.}{2014}]{Vera-Ciro2014ApJ}
{Vera-Ciro} C.,  {D'Onghia} E.,  {Navarro} J.,   {Abadi} M.,  2014, \mn@doi
  [\apj] {10.1088/0004-637X/794/2/173}, \href
  {https://ui.adsabs.harvard.edu/abs/2014ApJ...794..173V} {794, 173}

\bibitem[\protect\citeauthoryear{{Vera-Ciro}, {D'Onghia}  \&
  {Navarro}}{{Vera-Ciro} et~al.}{2016}]{Vera-Ciro2016}
{Vera-Ciro} C.,  {D'Onghia} E.,   {Navarro} J.~F.,  2016, \mn@doi [\apj]
  {10.3847/1538-4357/833/1/42}, \href
  {https://ui.adsabs.harvard.edu/abs/2016ApJ...833...42V} {833, 42}

\bibitem[\protect\citeauthoryear{{Villumsen}}{{Villumsen}}{1983}]{Villumsen1983ApJ...274..632V}
{Villumsen} J.~V.,  1983, \mn@doi [\apj] {10.1086/161475}, \href
  {https://ui.adsabs.harvard.edu/abs/1983ApJ...274..632V} {274, 632}

\bibitem[\protect\citeauthoryear{{Williams} \& {McKee}}{{Williams} \&
  {McKee}}{1997}]{Williams1997ApJ...476..166W}
{Williams} J.~P.,  {McKee} C.~F.,  1997, \mn@doi [\apj] {10.1086/303588}, \href
  {https://ui.adsabs.harvard.edu/abs/1997ApJ...476..166W} {476, 166}

\bibitem[\protect\citeauthoryear{{Wilson}, {Elmegreen}, {Bemis}  \&
  {Brunetti}}{{Wilson} et~al.}{2019}]{Wilson2019ApJ...882....5W}
{Wilson} C.~D.,  {Elmegreen} B.~G.,  {Bemis} A.,   {Brunetti} N.,  2019,
  \mn@doi [\apj] {10.3847/1538-4357/ab31f3}, \href
  {https://ui.adsabs.harvard.edu/abs/2019ApJ...882....5W} {882, 5}

\bibitem[\protect\citeauthoryear{{van der Kruit} \& {Searle}}{{van der Kruit}
  \& {Searle}}{1981}]{vander1981A&A....95..105V}
{van der Kruit} P.~C.,  {Searle} L.,  1981, \aap, \href
  {https://ui.adsabs.harvard.edu/abs/1981A&A....95..105V} {95, 105}

\makeatother
\end{thebibliography}








\bsp	
\label{lastpage}
\end{document}